\documentclass[openacc]{rsproca_new}
\usepackage{hyperref}

\titlehead{Research}

\begin{document}

\title{Wavenumber affects the lift of ray-inspired fins near a substrate}

\author{
Yuanhang Zhu$^{1,2,\dagger}$, Leo Liu$^{1}$, Tianjun Han$^{1,3}$, Qimin Feng$^4$, Keith W. Moored$^{3}$, Qiang Zhong$^4$, and Daniel B. Quinn$^{1,5,\ddagger}$}

\address{$^{1}$Department of Mechanical and Aerospace Engineering, University of Virginia, Charlottesville, VA 22904, USA\\
$^{2}$Department of Mechanical Engineering, University of California, Riverside, CA 92521, USA \\
$^{3}$Department of Mechanical Engineering and Mechanics, Lehigh University, Bethlehem, PA 18015, USA \\
$^{4}$Department of Mechanical Engineering, Iowa State University, Ames, IA 50011, USA\\
$^{5}$Department of Electrical and Computer Engineering, University of Virginia, Charlottesville, VA 22904, USA\\
$^{\dagger}$yuanhang.zhu@ucr.edu, $^{\ddagger}$danquinn@virginia.edu\\}

\begin{abstract}
Rays and skates tend to have different fin kinematics depending on their proximity to a ground plane such as the seafloor. Near the ground, rays tend to be more undulatory (high wavenumber), while far from the ground, rays tend to be more oscillatory (low wavenumber). It is unknown whether these differences are driven by hydrodynamics or other biological pressures. Here we show that near the ground, the time-averaged lift on a ray-like fin is highly dependent on wavenumber. We support our claims using a ray-inspired robotic rig that can produce oscillatory and undulatory motions on the same fin. Potential flow simulations reveal that lift is always negative because quasisteady forces overcome wake-induced forces. Three-dimensional flow measurements demonstrate that oscillatory wakes are more disrupted by the ground than undulatory wakes. All these effects lead to a suction force toward the ground that is stronger and more destabilizing for oscillatory fins than undulatory fins. Our results suggest that wavenumber plays a role in the near-ground dynamics of ray-like fins, particularly in terms of dorsoventral accelerations. The fact that lower wavenumber is linked with stronger suction forces offers a new way to interpret the depth-dependent kinematics of rays and ray-inspired robots.

\end{abstract}

\maketitle


\section{Introduction}\label{sec:intro}

\begin{figure}[ht!]
\centering
\includegraphics[width=0.9\textwidth]{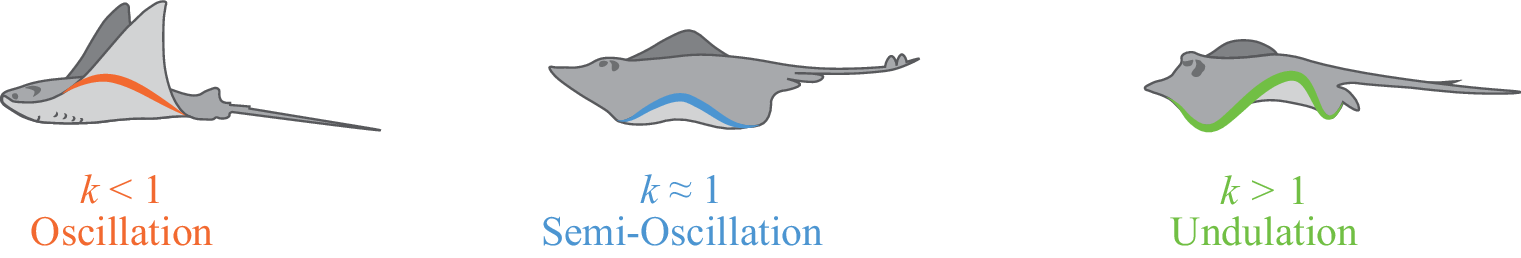}
\caption{Oscillatory, semi-oscillatory, and undulatory motions. Left: Cownose Ray (\emph{Rhinoptera bonasus}, $k=0.4$). Middle: Clearnose Skate (\emph{Rostroraja eglanteria}, $k=0.9$). Right: Bluespotted Ribbontail Ray (\emph{Taeniura lymma}, $k=1.4$). Wavenumbers extracted from \cite{rosenberger2001pectoral}.}
\label{fig.photo}
\end{figure}

Rays and skates swim differently depending on whether they are bottom-dwelling (benthic) or open-water-dwelling (pelagic). While bottom-dwelling rays tend to be more ``undulatory'' (high wavenumber), open-water-dwelling rays tend to be more ``oscillatory'' (low wavenumber) \cite{rosenberger2001pectoral,heine1992mechanics}. Here, the wavenumber is defined as the ratio between the ray's chord and wavelength. Motions in between ``undulatory'' and ``oscillatory'' have been called ``semi-oscillatory'' \cite{schaefer2005batoid} (Fig.~\ref{fig.photo}). Some rays use a mix of motions: smooth butterfly rays (\emph{Gymnura micrura}) tend to undulate near the seafloor but oscillate in open water \cite{rosenberger2001pectoral}. Existing explanations of this depth-dependent wavenumber focus on migration efficiency \cite{heine1992mechanics,blake2004fish} or sediment disturbance \cite{rosenberger2001pectoral}. We wondered if hydrodynamic interactions between fins and the seafloor could also play a role.

A ``ground'', such as the seafloor, induces a range of hydrodynamic effects on oscillating/undulating fins. These effects arise because proximity to a solid boundary modifies the surrounding flow field and pressure distribution, thereby altering the forces experienced by the fin. Rigid pitching hydrofoils (i.e., hydrofoils that pivot about their leading edge) produce more thrust near the ground \cite{quinn2014unsteady,dai2016self}, and they can experience positive or negative lift (i.e., pushing away or pulling toward the ground), depending on their distance from the ground \cite{kurt2019swimming}. These positive and negative lift forces result from the competition between positive wake-induced lift (arising from interactions between the hydrofoil and shed vortices as well as their ground-induced image vortices) and negative quasisteady lift (estimated from unsteady potential flow theory), while the added-mass lift (associated with fluid acceleration effects) remains zero for symmetric motions \cite{han2024revealing}. Moreover, the fin aspect ratio plays a critical role in determining the net lift:  the net lift of rigid pitching hydrofoils near the ground switches from positive to negative when the aspect ratio is decreased from 2.5 to 1 \cite{zhong2021aspect,han2024revealing}.

Oscillating/undulating flexible fins are more complicated. Like rigid fins, flexible fins can experience more thrust \cite{ren2022amplitude,fernandez2015large,park2017hydrodynamics,ryu2016flapping,zhang2017free,tang2016self,xie2020study} and both positive and negative lift \cite{ren2022amplitude,tang2016self,zhang2017free} near the substrate. However, a flexible fin's motion and its interaction with the substrate are highly coupled \cite{fernandez2015large,park2017hydrodynamics}, causing secondary effects. For a flexible fin, thrust may actually decrease near the substrate \cite{sierra2018two}, or it may only increase at high frequencies or resonance \cite{quinn2014flexible}. Regarding lift, some studies of near-ground flexible fins reported only negative lift and no equilibria \cite{shi2021ground,dai2016self,ren2022amplitude}. One study reported no significant near-ground changes at all \cite{blevins2013swimming}. With wavenumber as a passive output in these previous models--unlike in real rays and skates, where the wavenumber is actively modulated--it has remained difficult to isolate the apparently complex role of the oscillation-undulation spectrum for near-ground flexible swimmers.

The aim of the present study is to uncover the role of wavenumber in the near-ground dynamics of ray-like fins. We present here results from a robotic fin that can actively prescribe wavenumber, thereby isolating the effects of oscillation and undulation. We discovered that for near-ground fins, the primary effect of decreasing wavenumber is to increase the net suction toward the ground (increasing negative net lift). Far from the ground, both undulation and oscillation produce negligible net lift, because they are symmetric motions. Near the ground, oscillatory motions produce a suction force that is much larger than the force produced by undulatory motions, e.g.,~6 times larger at $f=3$ Hz (Fig.~\ref{fig.force_power}(\emph{f})). We used a vortex model to uncover the source of this suction force, which is an imbalance between wake-induced and quasisteady forces. Lastly, we employed a dynamical model to study the swimming and colliding behaviors of oscillatory and undulatory fins under the influence of the ground. Our results suggest that fin-vortex-ground interactions may have a role to play in the mystery of depth-dependent wavenumber.

\section{Methods}\label{sec:methods}

\begin{figure}
\centering
\includegraphics[width=1\textwidth]{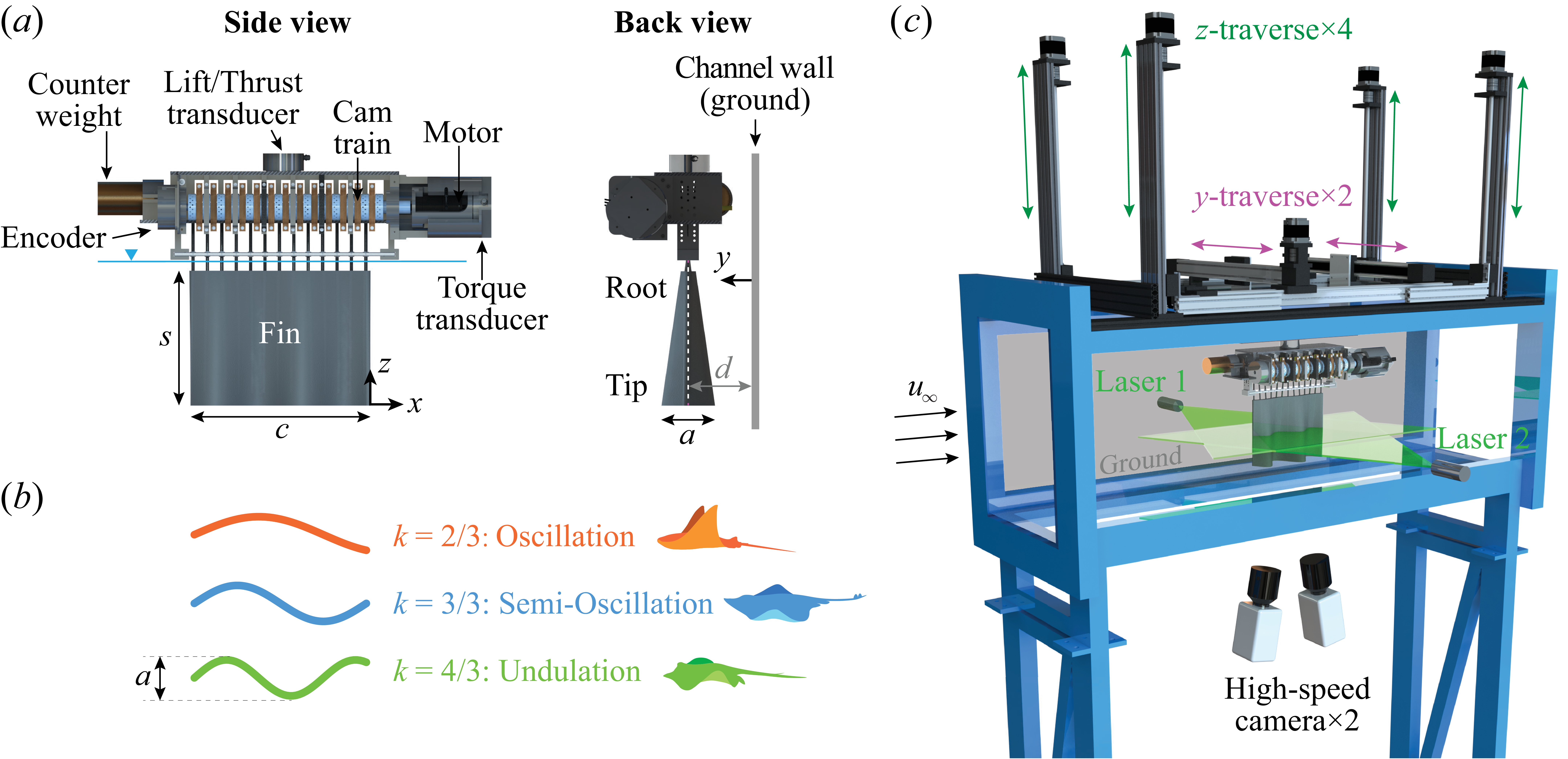}
\caption{(\emph{a}) Side view and back view of the ray-inspired robotic platform. (\emph{b}) Sketches of the wave shape at the fin tip. (\emph{c}) A perspective view of the experimental setup. Test section: 0.38$\times$0.45$\times$1.52 m).}
\label{fig.setup}
\end{figure}

To test the performance of our robotic fin, we measured the forces that it produced in a recirculating water channel. To better understand the origin of these forces, we used particle image velocimetry (experimental) and potential flow simulations (numerical). Lastly, we employed a dynamical systems model to study the swimming and colliding behaviors of the fin.

\subsection{Ray-inspired robotic fin}\label{sec:robot}

The robotic fin was molded using silicone rubber (50\% Ecoflex 0010 and 50\% Ecoflex 0030). It had a rectangular projected shape with a chord $c=270\;$mm, span $s=190\;$mm (aspect ratio $AR=s/c=0.7$), and thickness $h=9.5\;$mm. Undulatory and oscillatory motions were imposed by 13 garolite spines embedded in the fin (Fig.~\ref{fig.setup}(\emph{a})). The pivot point of the spines set the fin's peak-to-peak tip amplitude, $a$, to be 56$\;$mm. Each spine was connected to a rotating cam, and the offset between the cams dictated the fin's wavenumber, $k$, defined as the ratio between the fin's chord and wavelength. The modular cam train was driven by a geared motor (DJI M3508), which set the fin's frequency, $f$.

We tested kinematics that roughly align with real rays and skates. We tested three wavenumbers: $k=2/3$ (oscillation), $k=3/3$ (semi-oscillation), and $k=4/3$ (undulation) (Fig.~\ref{fig.setup}(\emph{b})). For comparison, rays in nature exhibit wavenumbers ranging from $\sim$ 0.4 to 1.4 \cite{rosenberger2001pectoral}. We chose frequencies corresponding to biologically-relevant Strouhal numbers, $St$. Like others before us \cite{clark2006thrust,dewey2012relationship}, we defined $St$ using the span-averaged peak-to-peak amplitude $A$ (i.e., the averaged peak-to-peak amplitude from root to tip, $A=a/2$): $St=fA/u_{\infty}=fa/2u_{\infty}$, where $u_{\infty}$ is the incoming flow speed. Here we used the averaged amplitude instead of the tip amplitude to calculate $St$ because the wake generated near the root of the fin is much smaller than that near the tip. Because amplitude varies linearly from root to tip, using this average amplitude accounts for the 2 we have in our denominator of $St$. Experimentally attainable frequencies varied by wavenumber, leading to $St$ ranges of $0.04-0.67$, $0.04-0.76$, and $0.04-0.86$ for the $k=2/3$, $3/3$, and $4/3$ fins, respectively. Further details of the fin assembly's design and construction can be found in \cite{zhong2022development}.

The robotic fin was suspended from a motorized 3-axis traverse into a recirculating water channel (Fig.~\ref{fig.setup}\emph{c}). We defined the fin's ground proximity, $d$, as the distance between the fin's mean position and the water channel's side wall (Fig.~\ref{fig.setup}\emph{a}). This proximity ranged from $d/c=0.27$ (or equivalently $d/a=1.3$, closest approach before contact) to 0.71 (or equivalently $d/a=3.4$, mid-channel position). Moving the two $y$-traverses together changes ground proximity $d$, while moving them differentially sets the fin's angle of attack with respect to the sidewall, $\alpha$. The channel's water level was held above the fin, and a horizontal baffle plate (not shown) was used to reduce surface waves. The free-stream speed of the channel was kept at $u_{\infty}=150\;\mathrm{mm~s^{-1}}$, monitored using an ultrasonic flow meter (Dynasonics TFXB). This speed corresponds to a chord-based Reynolds number $Re=\rho u_{\infty} c/\mu = 40,500$, where $\rho$ and $\mu$ are water density and dynamic viscosity.  

\subsection{Measuring fin forces and efficiency}\label{sec:forces}

To measure the forces produced by the fin, we used two six-axis force/torque transducers (ATI Mini40 IP65). One transducer (Calibration: SI-40-2) was mounted on top of the cam train to measure lift, $L$ (net $y$ force), and thrust, $T$ (negative net $x$ force). The other transducer (Calibration: SI-20-1) was mounted to the motor base to measure the motor's output torque, $\tau$. To balance the weight of the motor, we attached a brass counterweight on the upstream side of the cam train. An absolute encoder (US Digital A2K 4096 CPR) was attached to the end of the cam train to measure the angular position $\theta$ of the motor. The motor's output power was calculated as $P=\tau \dot{\theta}$. We report power measurements after subtracting power measurements taken in air so as to exclude frictional and internal-stress-induced power from the total power. Each combination of ground proximity and Strouhal number was repeated four times, and each trial was averaged over 30 cycles after a warm-up period of 5 cycles. Our force measurement data and the corresponding processing code are available at \cite{data}.

When reporting forces and power, we use the dynamic pressure to define dimensionless coefficients of lift ($C_L$), thrust ($C_T$), and power ($C_P$):

\begin{equation}
    \overline{C}_L = \frac{\overline{L}}{0.5\,\rho\,u_{\infty}^2\, s\,c},~\overline{C}_T = \frac{\overline{T}}{0.5\,\rho\,u_{\infty}^2\,s\,c},~\overline{C}_P = \frac{\overline{P}}{0.5\,\rho\,u_{\infty}^3\,s\,c},
\end{equation}

\noindent where overbars denote time-averaged quantities. We define the propulsive or ``Froude'' efficiency as $\eta\equiv \overline{C}_T/\overline{C}_P$, which estimates the fraction of the energy injected into the wake that is used for forward thrust.

\subsection{Measuring flow velocity fields}\label{sec:PIV}

We measured the three-dimensional velocity field around the fin using multi-layer stereoscopic particle image velocimetry (SPIV) (Fig.~\ref{fig.setup}\emph{c}). The flow was seeded with neutrally buoyant, silver-coated ceramic particles (50 $\mu m$ diameter, Potters Industries), and illuminated by two horizontal ($x$-$y$) laser sheets from opposing directions to reduce shadows ($532\,\text{nm}$, 5W Raypower MGL-W-532 and 10W CNI MGL-W-532A). Two high-speed cameras (Phantom SpeedSense M341, 2560$\times$1600 px) with $50\,\text{mm}$ lenses (Zeiss) imaged the illuminated particles from beneath the water channel. The cameras were outfitted with Scheimpflug adaptors (Dantec Dynamics) to manipulate the focus plane.

We used SPIV to measure the flow field at two wavenumbers and two ground proximities, for a total of four cases: ($k$, $d/c$) = ($2/3$, $0.71$); ($2/3$, $0.27$); ($4/3$, $0.71$); and ($4/3$, $0.27$). The fin has a constant Strouhal number $St=0.57$ for all four cases. For each case, we captured a $2.2\times1.5\times0.8c$ volume of three-dimensional three-component (3D3C) velocity vectors by stitching a total of 23 layers of 2D3C velocity vector fields together. For each layer, we captured 600 consecutive image pairs from each camera (15 cycles, 40 frames per cycle) at 120 Hz. The recorded image pairs were converted to two-dimensional three-component (2D3C) velocity fields in Dantec Dynamic Studio (v6.9) via an adaptive PIV algorithm (minimum interrogation window $32\times32$ px, maximum interrogation window $64\times64$ px). We then raised the entire fin by 1$\;$cm, using four synchronized $z$-traverses, and repeated the imaging process to complete another layer of SPIV scanning. We report time-averaged values (averaged across all 600 frames) and phase-averaged values (averaged across 15 oscillation periods in increments of 1/40th of a cycle). Our PIV data and the corresponding processing code are available at \cite{data}.

\subsection{Potential flow simulations}\label{sec:sim}

\begin{figure}
\centering
\includegraphics[width=0.6\textwidth]{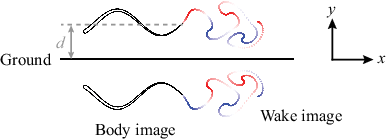}
\caption{A schematic of the potential flow simulation setup. An image system is used to enforce the no-flux boundary condition at the ground plane ($y=0$).}
\label{fig.sim_setup}
\end{figure}

To test the importance of fin aspect ratio and to better understand the origin of the suction force, we conducted two-dimensional potential flow simulations in parallel. In the simulations, we actuated a hydrofoil (3\% thickness, teardrop cross-section) near a ``ground'', implemented via the method of images (Fig. \ref{fig.sim_setup}). The method of images introduces body and wake images that replicate the ground's influence on the flow without explicitly modeling the ground itself. The free-stream speed ($u_\infty$), chord length ($c$), Strouhal number ($St=0.57$), and the wavenumbers ($k$) were matched to the experiments. Details of our algorithm can be found in prior work (e.g., Moored \cite{moored2018unsteady} and Han et al. \cite{han2024revealing}), but we reproduce the key components here.

Potential-flow simulations assume the flow is irrotational, incompressible, and inviscid, and therefore governed by Laplace's equation, $\nabla^{2}\Phi^{*}=0$, where $\Phi^{*}$ is the perturbation velocity potential in an inertial frame fixed to the undisturbed fluid \cite{lowspeed2001}. Two boundary conditions are imposed on $\Phi^{*}$: a no-flux condition at all surfaces ($\nabla \Phi^*\,\boldsymbol{\cdot\, n}=0$ where $\boldsymbol{n}$ is the surface normal vector) and a far-field decay condition ($\Phi^{*}\rightarrow 0$ at infinity).

The fin's surface, the wake vortex sheet, and their images are discretized into a finite number of boundary elements. Each body element contains a constant-strength doublet and source panel while each wake element contains a constant-strength doublet panel. In addition to the boundary conditions, an explicit Kutta condition is applied by setting the strength of the wake element at the trailing edge to enforce zero vorticity there. At each time step, a wake element is shed from the trailing edge with a strength that satisfies Kelvin's circulation theorem, and its strength remains constant thereafter. After being shed, wake elements advect with the fluid velocity. 

To compute the instantaneous pressure distribution on the fin surface, we apply the
unsteady Bernoulli equation in the inertial frame, which gives
\begin{equation}
p(x,y,t) \;=\;
-\rho\,\frac{\partial \Phi^*}{\partial t}\Big|_{\text{inertial}}
\;+\;\rho\,(\mathbf{u}_{\mathrm{rel}}+\mathbf{u}_\infty)\cdot\nabla \Phi^*
\;-\;\rho\frac{(\nabla \Phi^*)^2}{2}.
\label{eqn.bernoulli}
\end{equation}
Here $p$ denotes the unsteady pressure, and $\mathbf{u}_{\mathrm{rel}}$ is the local velocity
of the fin surface relative to the body-attached frame. Forces at each timestep are then obtained by integrating the pressure field over the fin surface,
\begin{equation}
\mathbf{F}(t) \;=\; \int_{S_b} -\,p(x,y,t)\,\mathbf{n}\,\mathrm{d}S ,
\label{eqn.force}
\end{equation}
where $S_b$ denotes the fin (body) surface, and $\mathbf{n}$ is the unit normal vector pointing into the fluid. The streamwise component of $\mathbf{F}$ gives the thrust $T$, and the wall-normal component gives the lift $L$. The corresponding power input $P$ is calculated as the surface integral of pressure times the local normal velocity of the fin:
\begin{equation}
P(t) \;=\; \int_{S_b} -\,p(x,y,t)\,\big(\mathbf{u}_{\text{rel}}\!\cdot\!\mathbf{n}\big)\, dS.
\label{eqn.power}
\end{equation}

To obtain further physical insights, we followed the approach of von Kármán \& Sears \cite{von1938airfoil}, where lift is decomposed into three components: added mass lift, wake-induced lift, and quasisteady lift. To calculate quasisteady and added mass lift, we reran the simulations with no wake elements shed. In these simulations, the unsteady term of the Bernoulli equation provides the added-mass pressure, and the steady term of the equation provides the quasisteady pressure. The wake-induced pressure is calculated by subtracting those two pressure components from the total pressure. See Han \emph{et al.} \cite{han2024revealing} for further details of our force decomposition method.

While Moored \cite{moored2018unsteady} presents a boundary element method capable of simulating both 2D and 3D swimmers, we deliberately chose the 2D version of this code to isolate the role of aspect ratio (i.e., the degree of three-dimensionality) on the ground effect observed in oscillatory and undulatory ray-like fins. This aspect ratio effect is important because as demonstrated in Zhong et al. \cite{zhong2021aspect} and Han et al. \cite{han2024revealing}, the net lift of rigid pitching hydrofoils near the ground switches from positive to negative when the aspect ratio is decreased from 2.5 to 1. The fins used in our experiments have a relatively low aspect ratio of approximately 0.7. By comparing the 3D experiments (finite aspect ratio) with 2D simulations (infinite aspect ratio), we were able to test if this aspect ratio effect on the net lift exists for  ray-like fins.

\subsection{Dynamical system models}\label{sec:dyn}

To study the swimming and colliding behaviors of oscillatory and undulatory fins under the influence of the ground, we created a simple dynamical system model based on our measured lift coefficients. In the model, we assumed neutrally buoyant point masses and calculated their trajectories by solving
\begin{equation}
    m \ddot{\vec{r}} = \left<0, \frac{1}{2}\,\rho\,u_0^2\,s\,c\,\overline{C}_L(y)\right>
\end{equation}
for $\vec{r}$, with initial conditions $\vec{r}=\left<0,y_0\right>$ and $\dot{\vec{r}}=\left<u_0, 0\right>$, where $m$ is the mass, $\vec{r}$ is the position vector and dots denote time derivatives. The parameters of our model were based on Southern Stingray (\emph{Dasyatis americana}), which has an averaged chord length (0.28m) close to our fin model (0.27m) \cite{rosenberger2001pectoral}. The mass was calculated to be $m=1.8$ kg using the disc width-weight relationship of the Southern Stingray reported in \cite{henningsen2010observations}, and the horizontal velocity $u_0$ was kept at 0.55 m/s to match the free-swimming velocity reported in \cite{rosenberger2001pectoral}.

\section{Results and discussions}\label{sec:result}
\subsection{Ground proximity affects lift more than thrust and power}\label{sec:ltp}

\begin{figure}
\centering
\includegraphics[width=1\textwidth]{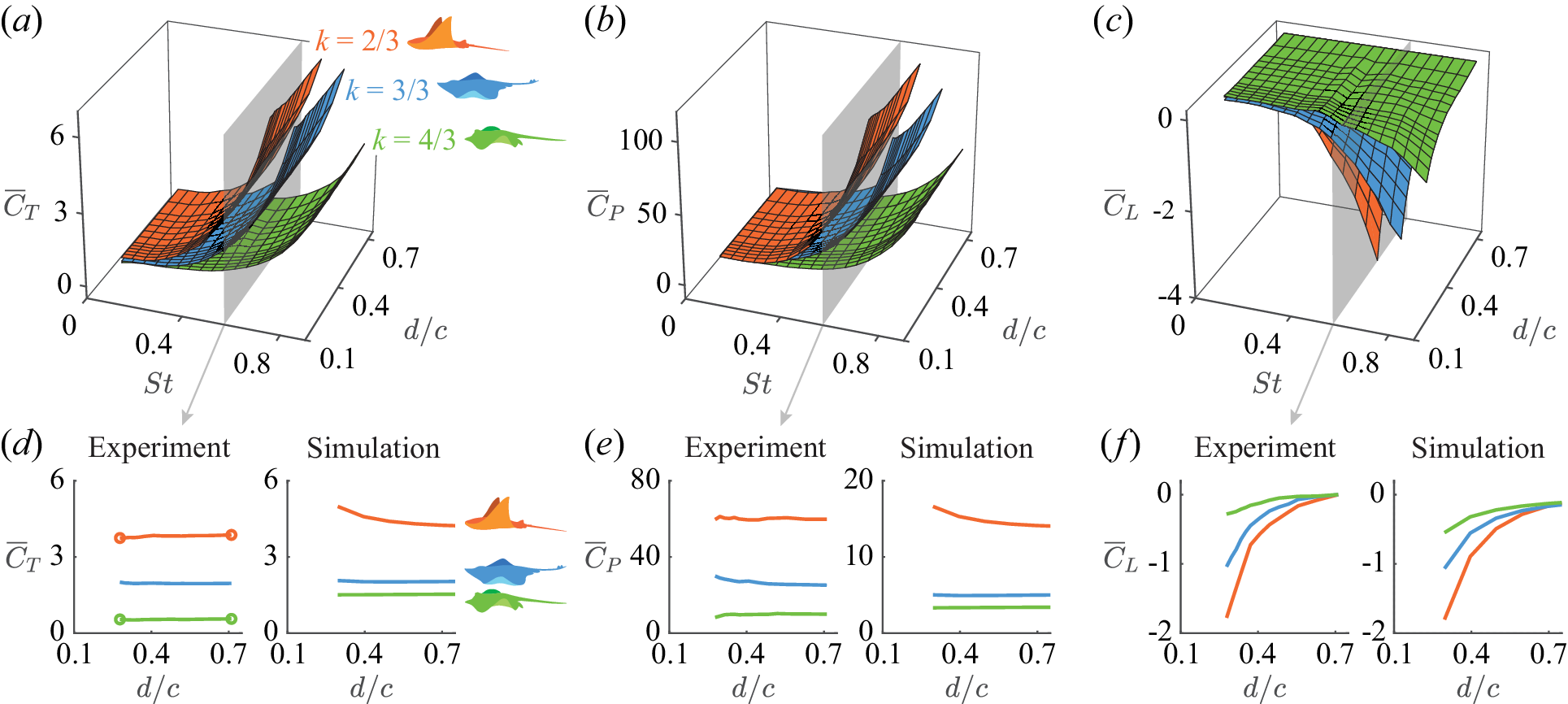}
\caption{(\emph{a-c}) Time-averaged thrust $\overline{C}_T$, power $\overline{C}_P$, and lift $\overline{C}_L$ coefficients of oscillatory ($k=2/3$), semi-oscillatory ($k=3/3$), and undulatory ($k=4/3$) fins at a range of Strouhal number $St$ and ground proximity $d/c$ ($d$: ground distance; $c$: chord length). Shaded boxes are slices through the 3D plots at $St=0.57$. (\emph{d-f}) Comparisons between experiments and simulations at $St=0.57$. Hollow circles in (\emph{d}) correspond to locations of PIV cases. Note the difference in the scale of $\overline{C}_P$ in (\emph{e}). The difference between experiment and simulation in $\overline{C}_P$ may be attributed to unmodeled physical effects in the 2D simulation.}
\label{fig.force_power}
\end{figure}

Far from the ground, thrust and power are functions of Strouhal number and wavenumber. With wavenumber fixed, higher Strouhal numbers produced more thrust and required more power, as they do for rigid foils \cite{wu2020review} (Fig.~\ref{fig.force_power}(\emph{a,b}), agreeing with Curet \emph{et al.} \cite{curet2011mechanical}). With Strouhal number fixed, lower wavenumbers produced more thrust and required more power, consistent with real rays, where oscillatory rays swim at lower frequencies than undulatory rays yet at comparable speeds \cite{rosenberger2001pectoral}. Across all wavenumbers and Strouhal numbers, the lift was approximately zero, as one would expect for (both spatially and temporally) symmetric motions far from a boundary (Fig.~\ref{fig.force_power}(\emph{c})). Also note that the fin shapes used in both our experiments and simulations were symmetric, unlike those of high-aspect-ratio pelagic rays which are generally of lift-producing shapes. Our experiments and simulations produced lift and thrust coefficients within 20$\%$ of each other (Fig.~\ref{fig.force_power}(\emph{d,f})), echoing previous findings that viscous forces play only a small role in near-ground dynamics \cite{quinn2014unsteady}. We attribute the larger differences in power coefficient (Fig.~\ref{fig.force_power}(\emph{e})) to unmodeled physical effects in the 2D simulation, including: (i) three-dimensional effects, where tip vortices, spanwise flow, and vortex stretching can alter the pressure distribution and therefore the power input; (ii) free-surface effects, which can modify the pressure on the fin body near the free surface and thus the power, even with the baffle plate in place; and (iii) channel blockage effects, which can change the velocity distribution near the fin and therefore the power, whereas the 2D simulation assumes an infinite domain.

Close to the ground, the measured thrust and power exhibited only minor effects. Thrust and power were unaffected by the ground in our experiments, and they showed a slight uptick at the closest ground proximities in our simulations (Fig.~\ref{fig.force_power}(\emph{d,e})). Our simulations, which are 2D, are consistent with prior experimental studies of 2D hydrofoils, where near-ground thrust showed moderate increases (e.g., a 44\% increase for $d/c=0.38$, $St=0.38$ \cite{quinn2014unsteady}). The absence of an observable thrust boost in our experiments is consistent with prior experiments \cite{blevins2013swimming} and simulations  \cite{shi2021ground} that have shown no near-ground changes in thrust for ray-inspired platforms. It could be that the low aspect ratios of ray-like fins preclude them from near-ground thrust benefits, as ground-induced forces are known to decrease with aspect ratio for rigid foils \cite{zhong2021aspect}.

Unlike thrust and power, the measured lift was significantly affected by ground proximity (Fig.~\ref{fig.force_power}(\emph{c,f})). This effect was most pronounced at high Strouhal numbers and low wavenumbers (high-frequency oscillatory motions). For example, at the lowest wavenumber ($k=2/3$), the lift coefficient fell from near zero at the channel centerline ($d/c=0.71$) to near -2 at the closest ground proximity ($d/c=0.27$) (Fig.~\ref{fig.force_power}(\emph{f})). Our results differ from studies of near-ground rigid foils, where both negative and positive lift were observed \cite{kurt2019swimming}, and our results support numerical studies of near-ground flexible heaving foils \cite{dai2016self} and undulatory fins \cite{shi2021ground,ren2022amplitude} where only negative (suction) lift was observed. The fact that our 2D simulation shows the same trend (Fig. \ref{fig.force_power}(\emph{f})) indicates it is not an aspect ratio effect. To better understand why lift is always negative, and why it is more negative for more oscillatory motions, we turned to the lift decomposition enabled by our potential flow model.

\subsection{Lift is negative because quasisteady lift overcomes wake-induced lift}\label{sec:lift}

\begin{figure}
\centering
\includegraphics[width=1\textwidth]{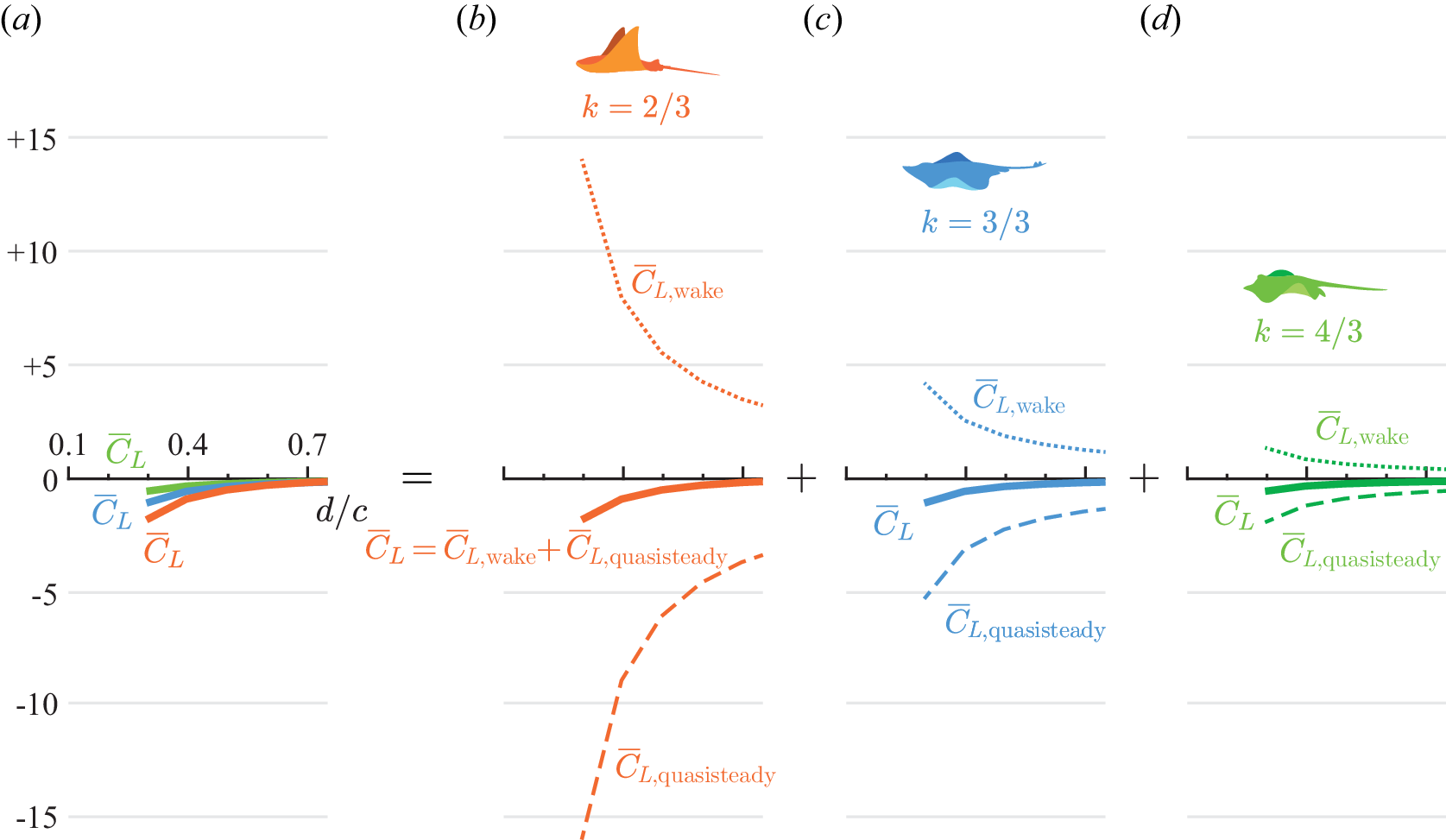}
\caption{Force decomposition results obtained from potential flow simulations. (\emph{a}) Total lift (\rule[0.5ex]{5mm}{1pt}) and its decomposition for (\emph{b}) oscillatory ($k=2/3$), (\emph{c}) semi-oscillatory ($k=3/3$), and (\emph{d}) undulatory ($k=4/3$) fins at $St=0.57$. The wake-induced lift $\overline{C}_{L,\mathrm{wake}}$ was represented by dotted lines (\rule[0.5ex]{0.4mm}{1pt}\hspace{0.4mm}\rule[0.5ex]{0.4mm}{1pt}\hspace{0.4mm}\rule[0.5ex]{0.4mm}{1pt}\hspace{0.4mm}\rule[0.5ex]{0.4mm}{1pt}\hspace{0.4mm}\rule[0.5ex]{0.4mm}{1pt}\hspace{0.4mm}\rule[0.5ex]{0.4mm}{1pt}) and the quasisteady lift $\overline{C}_{L,\mathrm{quasisteady}}$ was represented by dashed lines (\rule[0.5ex]{1.6mm}{1pt}\hspace{0.4mm}\rule[0.5ex]{1.6mm}{1pt}\hspace{0.4mm}\rule[0.5ex]{1.6mm}{1pt}).}
\label{fig.decompose}
\end{figure}

We decomposed time-averaged lift into added-mass, wake-induced, and quasisteady contributions: $\overline{C}_L = \overline{C}_{L,\mathrm{add}} + \overline{C}_{L,\mathrm{wake}} + \overline{C}_{L,\mathrm{quasisteady}}$. The time-averaged added-mass lift ($\overline{C}_{L,\mathrm{add}}$) was precisely zero, as it must be for the periodic motions considered here. The decrease in total lift near the ground (Fig.~$\ref{fig.decompose}(\emph{a})$) must therefore be explained by wake-induced and quasisteady lift only. We found that as the fin approached the ground ($d/c \rightarrow 0$), $\overline{C}_{L,\mathrm{wake}}$ rose sharply, while $\overline{C}_{L,\mathrm{quasisteady}}$ dropped sharply (Fig.~$\ref{fig.decompose}(\emph{b-d})$). Unlike previous work on rigid foils \cite{han2024revealing}, here the negative quasisteady lift always won out over the positive wake-induced lift, resulting in a negative total lift at all ground proximities. For insights into this effect, we looked to the origins of quasisteady and wake-induced lift.

Quasisteady lift is calculated without wake elements, so its magnitude is a function of the prescribed kinematics and effective flow velocity only \cite{Katz1985}. For rigid pitching fins, image vortices (i.e., fictitious vortices on the opposite side of the ground to satisfy the no-flux boundary condition, see also Section 2(\emph{d}) and Fig. \ref{fig.sim_setup}) reduce effective velocity on the downstroke more than they amplify effective velocity on the upstroke; the resulting asymmetry causes negative time-averaged quasisteady lift \cite{han2024revealing}. We suspect that a similar effect is taking place here for our lowest wavenumber fin (note that as $k\rightarrow 0$, the fin is effectively rigid). For example, for our lowest wavenumber fin at $St=0.57$, $\overline{C}_{L,\mathrm{quasisteady}}$ increased from $\sim$ 2 to 9 as $d/c$ dropped from $\sim$ 1 to 0.4 (Fig.~$\ref{fig.decompose}(\emph{b})$). In comparison, for rigid foils with $St=0.55$ \cite{han2024revealing}, $\overline{C}_{L,\mathrm{quasisteady}}$ increased from $\sim$ 2 to 5 over the same range. With increasing wavenumber, the image vortex system contains both positive and negative vortices, so its net influence on the effective velocity becomes weaker. This explains why the magnitude of $\overline{C}_{L,\mathrm{quasisteady}}$ decreases as the fin's wavenumber increases (Fig.~$\ref{fig.decompose}(\emph{c,d})$).

\begin{figure}
\centering
\includegraphics[width=1\textwidth]{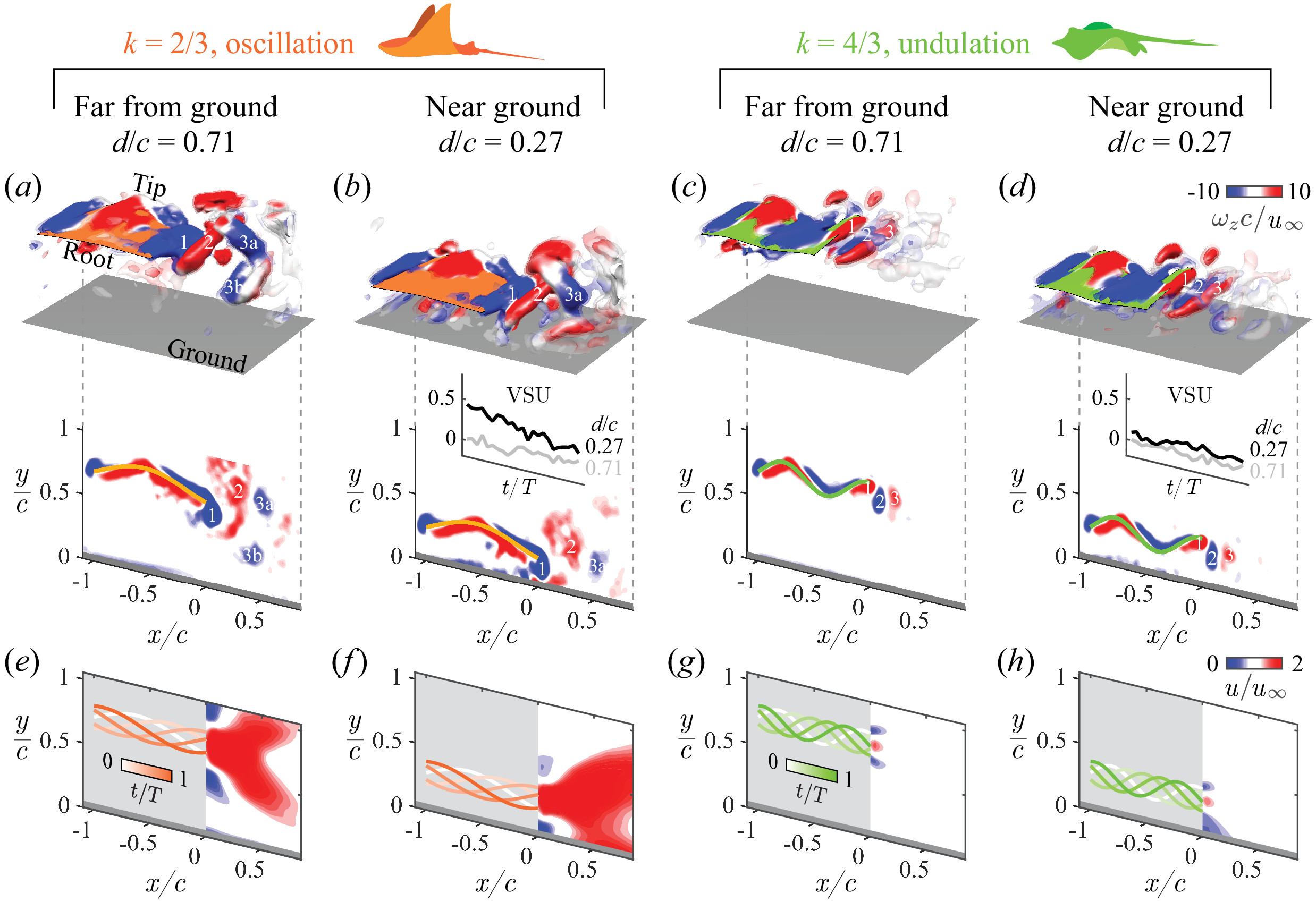}
\caption{(\emph{a-d}) Top: Phase-averaged 3D wake structures of oscillatory and undulatory fins far from the ground and near the ground at $St=0.57$, visualized by iso-$Q=10$ $s^{-2}$ surfaces ($Q$: second invariant of the velocity gradient tensor) and colored by the normalized spanwise vorticity $\omega_z c/u_{\infty}$. Bottom: Corresponding 2D vorticity plots sliced $0.11c$ above the fin tip. Insets: Vortex Spacing Unevenness (VSU) of near-ground ($d/c=0.27$, black) and far-from-ground ($d/c=0.71$, gray) cases. (\emph{e-h}) Time-averaged streamwise velocity normalized by the free-stream velocity, sliced at the same plane as the vorticity plots.}
\label{fig.vorticity}
\end{figure}

To understand the behavior of wake-induced lift, we turn to flow visualizations from our experiments (Fig.~\ref{fig.vorticity}). Wake-induced lift is caused by uneven spacing between vortices in the wake \cite{von2008piv}. We observed relatively even spacing in our wake when the fin is far from ground (Fig.~\ref{fig.vorticity}(\emph{b},\emph{d}) insets). When the fin is near the ground, we observe on average $\sim$ 30\% higher spacing between Vortices 1 and 2 than between Vortices 2 and 3a for the oscillatory fin (Fig.~\ref{fig.vorticity}(\emph{b}) inset), and $\sim$ 8\% higher spacing between Vortices 1 and 2 and Vortices 2 and 3 for the undulatory fin (Fig.~\ref{fig.vorticity}(\emph{d}) inset). Here we define Vortex Spacing Unevenness (VSU) as $(VS_{12}-VS_{23})/VS_{12}$, where $VS_{12}$ is the spacing between Vortices 1 and 2 and $VS_{23}$ is the spacing between Vortices 2 and 3. In contrast, differences as high as 700\% have been reported behind rigid foils \cite[see their figure 8(\emph{b})]{quinn2014unsteady}. Furthermore, Vortex 3 split into two pieces, 3a and 3b (Fig.~\ref{fig.vorticity}(\emph{a})), a phenomenon that has been linked to a reduction in wake-induced lift \cite{zhong2021aspect}. It appears that wake-induced lift was still present in our setup, but the wave-like motion of our fin did not produce enough vortex street unevenness to overcome quasisteady lift, as it does for rigid foils.

Wake-induced lift decreased with increasing wavenumber (Fig.~\ref{fig.decompose}(\emph{b-d})). This is consistent with our PIV measurements, where the ground compressed the wide momentum jet of the oscillatory fin while leaving the narrower jet of the undulatory fin largely unchanged (Fig.~\ref{fig.vorticity}(\emph{e-h})). Far from the ground, the oscillatory fin produced a bifurcating momentum jet, whereas the undulatory fin produced one weak jet (Fig.~\ref{fig.vorticity}(\emph{e,g})). Closer to the ground, one branch of the oscillatory fin's bifurcated wake was compressed by the wall, whereas the undulatory fin's wake was relatively unaffected (Fig.~\ref{fig.vorticity}(\emph{f,h})). A reduction in ground effect with increasing wavenumber is consistent with potential flow theory. As $k\rightarrow \infty$, the flow would need to be entirely horizontal to satisfy the no flux condition at the fin's surface; in this extreme, an image vortex system is no longer needed to satisfy no flux at the ground, i.e.,~the ground would have no effect on the fin.

\subsection{Implications for biological and engineered systems}\label{sec:implication}

For the ray-like motions considered here, ground proximity did not significantly affect thrust and power (Fig. \ref{fig.force_power}(\emph{d,e})). Ground proximity therefore did not significantly affect the propulsive efficiency ($\eta\equiv\overline{C}_T/\overline{C}_P$). Indeed, across all the cases we considered, efficiency changed by an average of only $\pm$0.82\% between the water channel midline and the closest ground proximity. Ground effects aside, wavenumber did affect efficiency: peak efficiency occurred around $St=0.3$ when $k=4/3$ and $St=0.6$ when $k=2/3$, but these trends are the same regardless of ground proximity. Differences in efficiency may help to explain why high wavenumber rays use different kinematics than low wavenumber rays---for example, higher fin-beat frequencies \cite{heine1992mechanics}---but they offer no guidance for why certain wavenumbers may be better near the ground.

\begin{figure}
\centering
\includegraphics[width=0.9\textwidth]{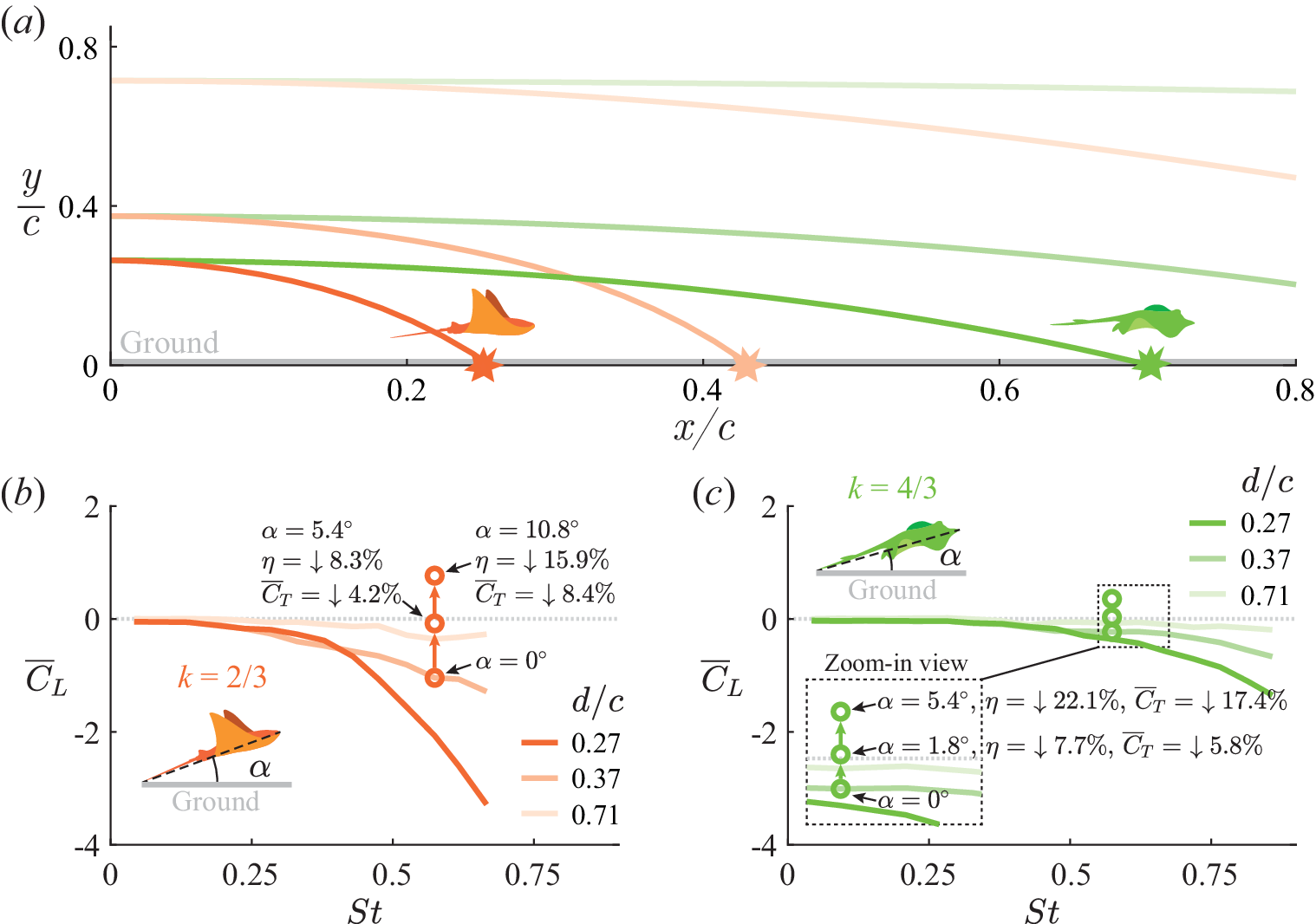}
\caption{(\emph{a}) Swimming and colliding of oscillatory ($k=2/3$) and undulatory ($k=4/3$) fins released at different ground distances, simulated by using experimental data at $St=0.57$. (\emph{b,c}) How changing the angle-of-attack affects the lift coefficient.}
\label{fig.AoA}
\end{figure}

Wavenumber did, however, have a large effect on time-averaged lift near the ground. To illustrate the magnitude of this effect, we constructed a simple dynamical system model based on our measured lift coefficients using the method described in Section 2(\emph{e}). Here we modeled the fin as point masses and because of the suction force, the masses quickly collide with the ground, with the oscillatory fin colliding the quickest and the undulatory fin colliding the slowest (Fig.~\ref{fig.AoA}(\emph{a})). These results are consistent with behaviors observed in real rays and skates, as undulating species often coast close to the substrate.

To avoid these suction-driven collisions (e.g., oscillatory cownose rays foraging near the substrate \cite{ajemian2013foraging}) would require asymmetric kinematics to offset the suction with upward lift. The asymmetry could be a faster downstroke or a pitch bias---both strategies that have been observed in real rays \cite{blevins2012rajiform, fish2016hydrodynamic,ren2022amplitude}. To test this idea, we repeated our force and power measurements for the oscillatory and undulatory fins with different angles of attack $\alpha$ at a fixed $St=0.57$ and $d/c=0.37$ measured from the mid chord (Fig.~\ref{fig.AoA}(\emph{b,c})). To achieve level swimming, our oscillatory fin would need $\alpha \approx 5.4^\circ$, while our undulatory fin would only need $\alpha \approx 1.8^\circ$. These angles are comparable to those observed in live rays \cite{blevins2012rajiform}. If a ray or ray-like robot needed to offset negative lift with a pitch bias, the required pitch bias would be lower for an undulatory fin. Increasing $\alpha$ leads to lower efficiency and thrust (Fig.~\ref{fig.AoA}(\emph{b,c})), presumably due to an increase in body drag. More detailed investigations into the effect of $\alpha$ on the variation of $\eta$ and $\overline{C}_T$ across different wavenumbers could provide valuable insights for the development of control strategies. However, such analysis is beyond the scope of the present study and is recommended for future work.

In addition to near-ground foraging, the results presented above may have potential relevance to the burial mechanics of benthic fishes \cite{seamone2021ocellate,ren2024numerical}, extending the biological implications of our findings. The suction forces and fin-vortex-ground interactions identified in our study may play a role in sand fluidization, a key mechanism used by flounders and rays to bury themselves for predator avoidance or ambush predation. Furthermore, these insights could inform the design of next-generation autonomous underwater vehicles (AUVs) that need to traverse the seafloor or maintain position beneath the sand for stealth tasks.

One potentially important factor affecting the dynamics of ray-like motions near the substrate--but not considered in the present study--is the oscillation of the center of mass due to inertia. In real fish (or untethered vehicles) swimming close to the substrate, the center of mass oscillates vertically as the fins move. This effect could influence how ground proximity or Strouhal number affects the swimming performance, similar to observations in previous studies \cite{wen2013understanding,das2022contrasting}. In our experiments, all force and power measurements were time-averaged, and instantaneous variations in these quantities were not characterized. The influence of inertial effects on ray-like motion is challenging to assess with a robotic platform like ours, but could potentially be explored using cyber-physical systems \cite{mackowski2011developing,zhu2020nonlinear} or numerical simulations.

Another potentially important factor for ray-like swimming near the substrate is the shape of the fin platform. In the present study, we used a rectangular fin, but real rays have different fin shapes (e.g., cownose rays have triangular fins, and bluespotted ribbontail rays have semi-circular fins), which may alter fin-wake-ground interactions and, consequently, the hydrodynamic forces. While these factors are beyond the scope of the present study, we acknowledge their importance and encourage future investigations to address them. We also acknowledge that future work should extend our present 2D potential flow simulations to 3D to capture the full complexity of ground effect in flexible, ray-like fins, especially where tip vortices or spanwise flow may play a role.

\section{Conclusions}\label{sec:conclusion}

We have shown that ray-like fins produce a net suction force toward the ground when actuated with symmetric oscillations. This suction force has been observed in prior simulations \cite{dai2016self,shi2021ground,ren2022amplitude}; here we offer the first experimental confirmation and decomposition analysis of this force. Our analysis reveals that oscillatory motions (lower wavenumbers) lead to higher suction forces. If it were disadvantageous to have a net suction force near the ground, more undulatory motions (high wavenumber) may be desirable. While our results cannot prove whether or how wavenumber affects locomotion and morphology in bottom-dwelling rays, they do suggest that wavenumber's effect on net lift could be a contributing factor.

\ack{This work was supported by NSF (Grant 2040351, Program Manager: Ron Joslin) and ONR MURI (Grant N00014-22-1-2616, Program Manager: Bob Brizzolara).}


\vskip2pc

\bibliographystyle{RS}

\bibliography{ref}

\end{document}